\documentclass[a4paper,11pt]{article}
\pdfoutput=1
\usepackage{jcappub} 
 \usepackage{ulem}
\usepackage[lofdepth,lotdepth,caption=false]{subfig}
\usepackage[T1]{fontenc}

\usepackage{amsmath}

\title{Gamma-ray bounds from EAS detectors and heavy  decaying dark matter constraints}

\author[a]{Arman Esmaili}
\author[b]{and Pasquale Dario Serpico}

\affiliation[a]{INFN, Laboratori Nazionali del Gran Sasso, Assergi (AQ), Italy}
\affiliation[b]{LAPTh, Univ. de Savoie Mont Blanc, CNRS, B.P.110, Annecy-le-Vieux F-74941, France}

\emailAdd{arman.esmaili@lngs.infn.it}
\emailAdd{serpico@lapth.cnrs.fr}

\subheader{LAPTH-026/15}
\abstract{The very high energy Galactic $\gamma$-ray sky is partially opaque in the ($0.1-10$) PeV energy range. In the light of the recently detected high energy neutrino flux by IceCube, a comparable very high energy $\gamma$-ray flux is expected in any scenario with a sizable Galactic contribution to the neutrino flux. Here we elaborate on the peculiar energy and anisotropy features imposed upon these very high energy $\gamma$-rays by the absorption on the cosmic microwave background photons and Galactic interstellar light. As a notable application of our considerations, we study the prospects of probing the PeV-scale decaying DM scenario, proposed as a possible source of IceCube neutrinos, by extensive air shower (EAS) cosmic ray experiments. In particular, we show that anisotropy measurements at EAS experiments are already sensitive to $\tau_{\rm DM}\sim {\cal O}(10^{27})$~s  and future measurements, using better gamma/hadron separation, can improve the limit significantly.}

\begin{document}
\maketitle
\flushbottom

\section{Introduction}
\label{sec:intro}

The very high energy (VHE) $\gamma$-ray band, loosely defined as the one above ${\cal O}$(100)$\,$GeV, is peculiar in many respects. For instance, on the detection side, it is at the upper edge of the energy range which can be probed from space via direct techniques, and can be effectively studied only by ground-based telescopes. An important theoretical feature in the study of VHE $\gamma$-rays is that {\it absorption is relevant}, i.e. the {\it extragalactic} sky is not transparent to VHE $\gamma$-rays due to the pair-production absorption onto the Extragalactic Background Light (EBL) and the cosmic microwave background (CMB). Actually, for photon energies of ${\cal O}(100)$~TeV even the Galactic diffuse radiation field starts playing a role, while at $\sim$ PeV energies the mean-free path of $\gamma$-rays due to the absorption on CMB photon bath approaches ${\cal O}(10)$~kpc, comparable to the distance to Galactic Center (GC). Hence, for VHE $\gamma$-rays even the Galactic sky becomes partially opaque. While the opacity of the extragalactic sky is routinely taken into account in the analysis of cosmologically distant active galactic nuclei, both by the Fermi-LAT space telescope~\cite{Ackermann:2012sza} and by the ground based instruments, see e.g.~\cite{Abramowski:2012ry}, the absorption at Galactic scale is usually neglected, although the theoretical importance of the phenomenon has been acknowledged in the past, see e.g.~\cite{Moskalenko:2005ng}.

The most obvious reason for the relative lack of interest in this phenomenon is that the ``PeVatron'' window in $\gamma$-ray astrophysics has yet to be opened, with current telescopes running out of statistics below ${\cal O}$(100) TeV (even for the most powerful sources) to be sensitive to spectral distortions. Another reason may simply be that PeVatrons (in gamma-rays) in our Galaxy are expected to be rare, if present at all, requiring extreme acceleration parameters. The situation has however changed in the last couple of years, following the discovery by the IceCube collaboration of a diffuse neutrino flux in the 30 TeV-2 PeV range~\cite{Aartsen:2013bka,Aartsen:2013jdh,Aartsen:2014gkd}. Virtually in any conceivable model for its origin, one expects an associated $\gamma$-ray flux with similar energy budget in the $\sim$~PeV energy range. The main unknown concerns the distance at which the sources of these events are located: if the neutrino flux is due to a population at cosmological distances, the absorption should be so severe that the initial VHE flux would cascade down below ${\cal O}$(100)$\,$GeV, contributing to the residual isotropic background precisely measured by Fermi~\cite{Ackermann:2014usa}. However, if it is due to Galactic or nearby extragalactic sources (see e.g.~\cite{Razzaque:2013uoa,Lunardini:2013gva,Lunardini:2015laa,Ahlers:2015moa}), the associated gamma-ray flux should still be detectable at VHE, albeit with a significantly suppressed spectrum. This has been noted soon after the discovery, see for instance~\cite{Ahlers:2013xia} or~\cite{Supanitsky:2013ooa}, and current $\gamma$-ray constraints are one argument disfavoring close-by discrete Galactic sources for the majority of the events. 

The calculation of both the expected signal and the observational constraints is however more involved in the case of truly diffuse local sources associated with the IceCube data, such as in an astrophysical origin in the Galactic halo~\cite{Taylor:2014hya}, or a decaying Dark Matter (DDM) origin, see~\cite{Esmaili:2013gha,Esmaili:2014rma,Bhattacharya:2014vwa,Rott:2014kfa,Fong:2014bsa} (see also~\cite{Zavala:2014dla,Bhattacharya:2014yha,Kopp:2015bfa,Daikoku:2015vsa} for some DM related interpretations of the IceCube data). On the one hand, the expected signal is usually {\it anisotropic}, at very least due to our off-set position with respect to the center of the Galactic Halo, so its detailed calculation requires at least a 2D modeling of the problem, and multiple numerical integrations, a complication that typically is not fully taken into account even in the most recent calculations~\cite{Murase:2015gea}. On the other hand the observational bounds (the most constraining ones being the one published by CASA-MIA~\cite{Chantell:1997gs} and those reported in a proceeding by KASCADE~\cite{KASCADE03}) are derived: {\it a}) based on observations of a limited portion of the sky; {\it b}) typically assuming a isotropic $\gamma$-ray flux.

The hypothesis {\it b}) is clearly incorrect, since even an incoming isotropic flux would acquire an anisotropy due to the anisotropic absorption. The hypothesis {\it a}) means that a proper use of these constraints would require to re-run ad hoc analyses by the collaborations, based on specific {\it energy and angular-dependent} templates, to be convoluted with the detector characteristics. This is a pity, since as we have argued in~\cite{Esmaili:2014rma}, for an important class of scenarios like the DDM ones, this kind of data are what currently comes the closest to an independent test of the hypothesis~\footnote{Note that a superficial look at Fermi-LAT isotropic gamma-ray background (IGRB) results might suggest that they are already very constraining, notably thanks to the few high-energy points. We stress here that they are unfortunately not robust with respect to the IGRB extraction procedure from the Extragalactic gamma-ray background (EGB). In more technical terms, the determination of the IGRB by the Fermi-LAT team does not take into account the uncertainty in the subtraction of the point sources contribution (very uncertain at high energies, relying on an extrapolation), which would constitute the dominant source of uncertainty in the last IGRB points. If the EGB is conservatively used, instead, the bounds are degraded, as can be seen in~\cite{Murase:2015gea}.}.

Although the important role of EAS probes of this scenario has been discussed in the past (see e.g.~\cite{Ahlers:2013xia,Esmaili:2014rma,Murase:2015gea}), here we revisit the calculation of the expected $\gamma$-ray flux, with a triple goal: {\it i}) To estimate more precisely the spectral and angular shape of a DDM signal, with state of the art treatment for the primary $\gamma$-ray absorption and the inverse Compton component. {\it ii}) To point out that due to the generically anisotropic nature of the VHE $\gamma$-ray component, even detectors with little or without gamma-hadron rejection capability should be able to put constraints on these contributions based merely on the expected {\it anisotropy}. {\it iii}) To motivate experimental collaborations to specifically constrain some angular-energy templates, to optimize their constraining power for specific models. For the DDM case, for instance, an intermediate step in this direction would be to derive (energy-dependent) bounds in coronas around the GC, in Galactic cylindrical coordinates. We also discuss the greatly improved potential of detectors with significant hadron vs. $\gamma$-ray  rejection capabilities.

As a case study we consider throughout this paper the peculiarities of the expected $\gamma$-ray flux from DDM. Yet, similar considerations would apply to any other Galactic diffuse flux model (a few examples have been listed e.g. in~\cite{Kalashev:2014vra}). We will consider both the prompt $\gamma$-ray flux and the flux from inverse Compton (IC) scattering of $e^\pm$ off the ambient photon bath. Both contributions would be present also in other diffuse flux models: in the commonly considered astrophysical hadronic production of neutrinos, they are associated to prompt $\gamma$-rays from $\pi^0$ decay and $e^\pm$ from $\pi^\pm$ decay, respectively. The specific features studied in this paper, mainly the inherent anisotropy due to absorption at Galactic scale and the peculiar profile of IC flux due to diffusion/losses of PeV $e^\pm$,  would similarly provide powerful diagnostic tools in probing alternative diffuse flux models. The only differences would be in the initial spectra and the geometric distribution of the source term in the Galaxy/Galactic halo.

This article is structured as follows: in section~\ref{sec:absorption} we describe the peculiar energy-angular dependence of the $\gamma$-ray flux absorption, and our computation of the $\gamma$-ray opacity. In section~\ref{sec:constraints} we compare the expected $\gamma$-ray flux from DDM with current constraints from EAS experiments, as well as the diagnostic power of forthcoming experiments. The two components of the $\gamma$-ray flux from DDM, prompt and IC flux, are discussed in sections~\ref{sec:prompt} and \ref{sec:ICflux}, respectively. Section~\ref{sec:aniso} is devoted to the discussion of the expected anisotropy of the total cosmic ray flux. Finally, in section~\ref{sec:conc} we conclude.

\section{Absorption of $\gamma$-rays at Galactic scale}
\label{sec:absorption}

The $\gamma$-ray flux in the approximate range $10^{-2}\div10^2$~PeV will suffer attenuation in the Galaxy due to the pair production $\gamma\gamma\to e^-e^+$ process onto photon baths: at the lower energies, starlight (SL) and infrared (IR) photons constitute important targets (mostly however for directions towards the inner Galaxy), while at $\sim$~PeV energies and above the homogeneous cosmic microwave background (CMB) is dominant. In the following we calculate the optical depth $\tau_{\gamma\gamma}$ for both CMB and SL+IR, for different incoming directions and energies. 

\begin{figure}[th]
\centering
\subfloat[]{
\includegraphics[width=0.5\textwidth]{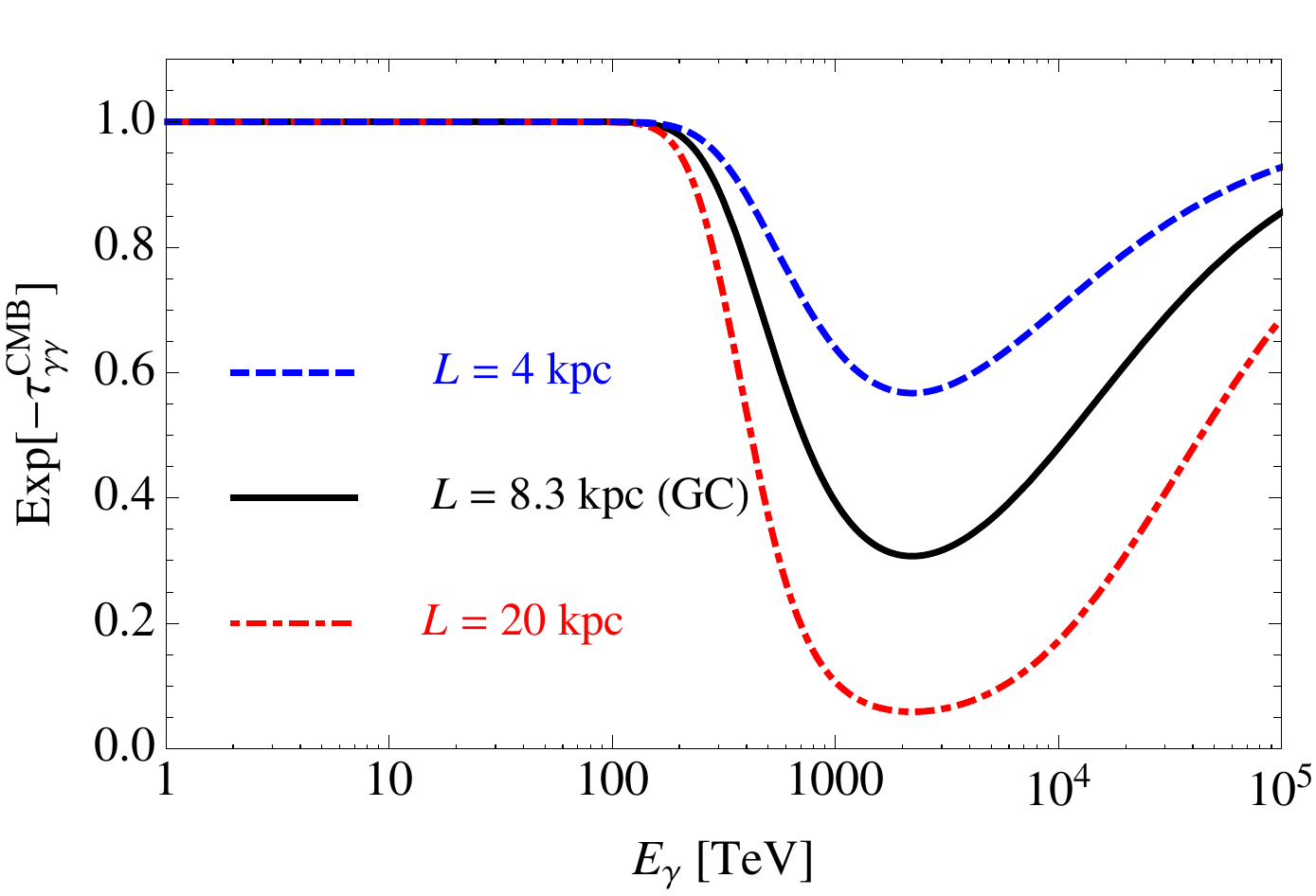}
\label{fig:cmb-abs-1D}
}
\subfloat[]{
\includegraphics[width=0.5\textwidth]{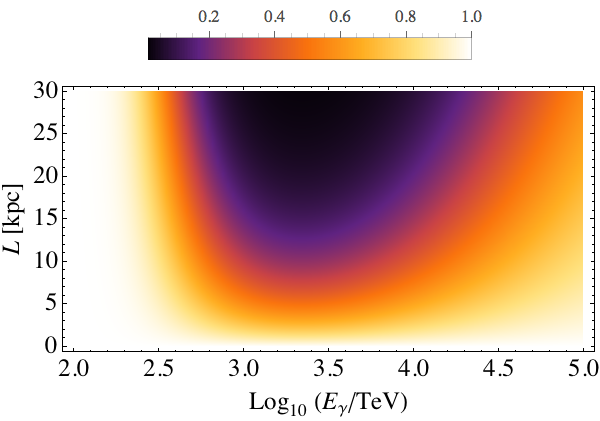}
\label{fig:cmb-abs-2D}
}
\caption{\label{fig:cmb-abs} Plot of the absorption of $\gamma$-rays on CMB photons. Panel (a): for a source at distance $L=4$~kpc, 8.3 kpc and 20 kpc. Panel (b): 2D density plot of $\exp[-\tau_{\gamma\gamma}^{\rm CMB}]$ as function of $L$ and $E_\gamma$.}
\end{figure}

For the technically simpler case of pair production on CMB photons, the optical depth for photons of energy $E_\gamma$ coming from a source at distance $L$ can be calculated as (here and in the following, we use natural units with $c=k_B=1$) 
\begin{equation}\label{eq:cmb-abs}
\tau_{\gamma\gamma}^{\rm CMB} (E_\gamma,L) = L \iint \sigma_{\gamma\gamma} (E_\gamma,\varepsilon)\, n_{\rm CMB}(\varepsilon) \, \frac{1-\cos\theta}{2} \sin\theta{\rm d}\theta\,{\rm d}\varepsilon~,
\end{equation}
where $\sigma_{\gamma\gamma}$ is the pair production cross section given by
\begin{equation}
\sigma_{\gamma\gamma} = \frac{\pi}{2}\frac {\alpha^2}{m_e^2} (1-\beta^2) \left[ (3-\beta^4) \ln \left( \frac{1+\beta}{1-\beta} \right) - 2\beta(2-\beta^2) \right]~,
\end{equation}
where 
\begin{equation}
\beta = \sqrt{1-1/s}\quad , \; {\rm and}  \quad s=\frac{\varepsilon E_\gamma}{2m_e^2} (1-\cos\theta)\,,
\end{equation}
 $\alpha$ is the fine-structure constant, $m_e$ the electron mass and $\theta$ is the angle between the momenta of photons. The $n_{\rm CMB}(\varepsilon)$ is the differential number density of CMB photons given by
\begin{equation}
n_{\rm CMB}(\varepsilon) = \frac{\varepsilon^2}{\pi^2} \frac{1}{e^{\varepsilon/T_{\rm CMB}}-1}~,
\end{equation}
where $T_{\rm CMB}=2.348\times10^{-4}$~eV. By changing variable $\varepsilon\to\varepsilon_c$, where $\varepsilon_c=\sqrt{\varepsilon E_\gamma (1-\cos\theta)/2}$ is the photon center of momentum energy, and performing the integral on $\theta$, the expression for $\tau_{\gamma\gamma}^{\rm CMB}$ can be reduced to a single integral to be performed numerically, namely 
\begin{equation}
\tau_{\gamma\gamma}^{\rm CMB} (E_\gamma,L) = \frac{-4T_{\rm CMB}L}{\pi^2E_\gamma^2} \int_{m_e}^{\infty}  \varepsilon_c^3\, \sigma_{\gamma\gamma} (\varepsilon_c) \ln\left[ 1- e^{-\frac{\varepsilon_c^2}{E_\gamma T_{\rm CMB}}} \right] {\rm d}\varepsilon_{c}~.
\end{equation}
Figure~\ref{fig:cmb-abs-1D} shows the $\tau_{\gamma\gamma}^{\rm CMB}$ as function of $E_\gamma$ for three different values of $L=4$~kpc, 8.3 kpc and 20 kpc. As can be seen, for a source of $\gamma$-ray at Galactic center (GC), at about $L=8.3$~kpc, the absorption is $\sim70\%$ at $E_\gamma\sim2$~PeV. Figure~\ref{fig:cmb-abs-2D} shows the contour plot of $\exp[-\tau_{\gamma\gamma}^{\rm CMB}]$, as function of photon energy $E_\gamma$ and source distance $L$. 

\begin{figure}[!t] 
\centering
\includegraphics[width=0.85\textwidth]{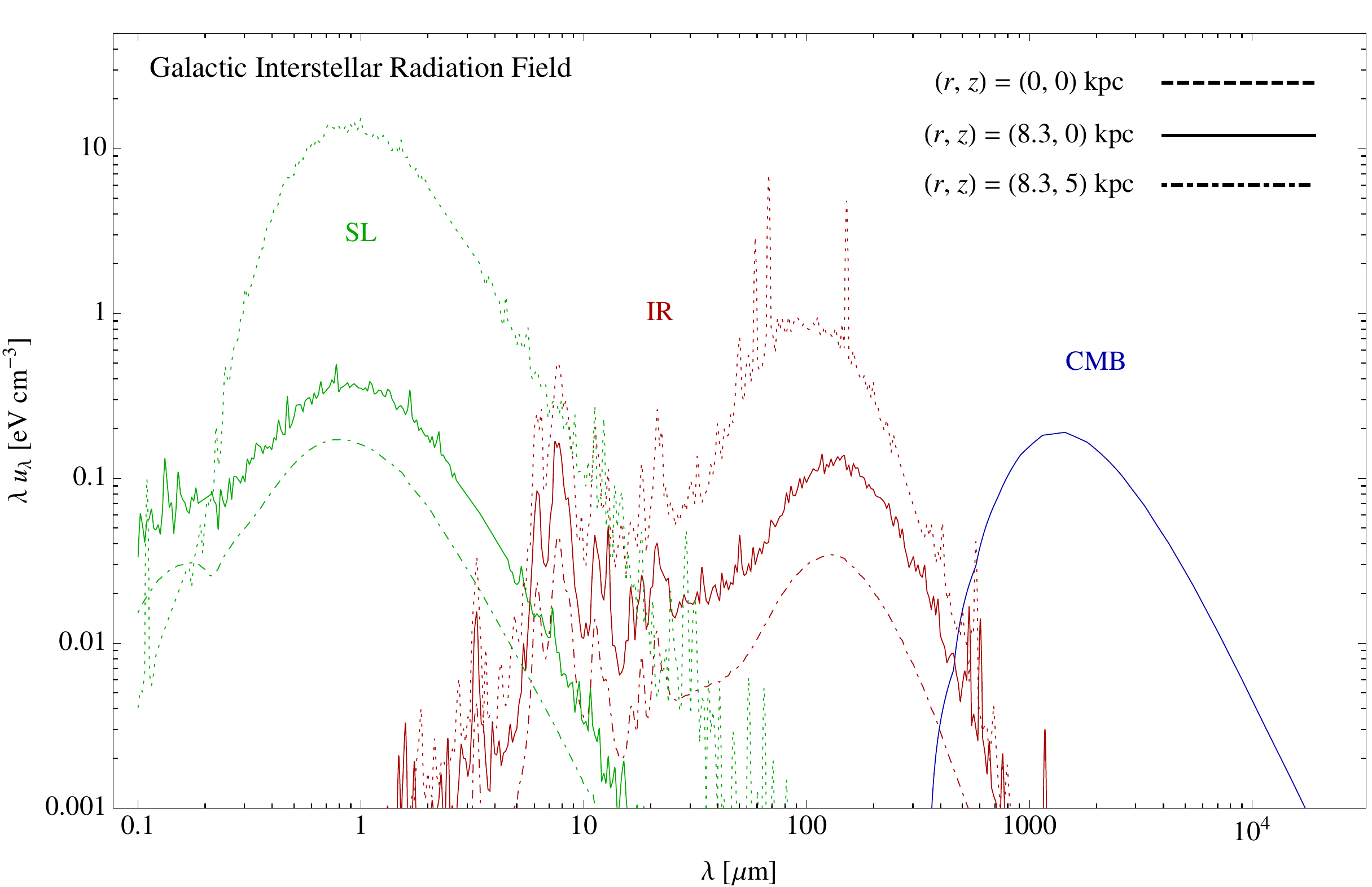} 
\caption{The energy density of ISRF (including starlight, IR and CMB), extracted from GALPROP~\cite{galprop},  at three different positions in our Galaxy: the dotted curves are for $(r,z)=(0,0)$, that is GC; the solid curves are for $(r,z)=(8.3,0)$~kpc, that is the Sun position, and the dot-dashed curves are for $(r,z)=(8.3,5)$~kpc.}
   \label{fig:nslir}
\end{figure}

The optical depth due to pair production on the SL+IR photon bath can be calculated similarly to eq.~(\ref{eq:cmb-abs}), with the extra complication that  the integral along the line of sight is non-trivial, since the photon bath number density $n_{\rm SL+IR}$  also depends on position {\bf x}, and the optical depth also depends on the Galactic coordinates $(b,l)$. In the approximation that the photon field is inhomogeneous but isotropic one can write
\begin{equation}\label{eq:sl-abs}
\tau_{\gamma\gamma}^{\rm SL+IR} (E_\gamma,L,b,l) =\int_{0}^{L} {\rm d}s \iint \sigma_{\gamma\gamma} (E_\gamma,\varepsilon) \, n_{\rm SL+IR}\left[\varepsilon, {\bf x}(s,b,l)\right]  \frac{1-\cos\theta}{2} \sin\theta{\rm d}\theta\,{\rm d}\varepsilon~,
\end{equation}
where the line-of-sight parameter $s$ is related e.g. to the cylindrical coordinates $(r,z)$, with the origin at the GC, by
\begin{equation}
r = \sqrt{R_\odot^2 + s^2 \cos^2b - 2 s R_\odot \cos b \cos l} \quad {\rm and} \quad z=s\sin b~,
\end{equation}
where $R_\odot\simeq 8.3\,{\rm kpc}$ is the distance of the Sun to the GC. The number densities of SL and IR photons have been extracted from the GALPROP code~\cite{galprop} and their energy densities for some representative positions are plotted in figure~\ref{fig:nslir}. Obviously, the CMB radiation field is homogenous and thus pervades the whole Galaxy uniformly, while the SL and IR components of radiation field are clearly position dependent: larger at GC and in the Galactic disk, decreasing rapidly by moving perpendicularly from Galactic disk, along the $z$ direction.  

\begin{figure}[!t] 
\centering
\includegraphics[width=0.7\textwidth]{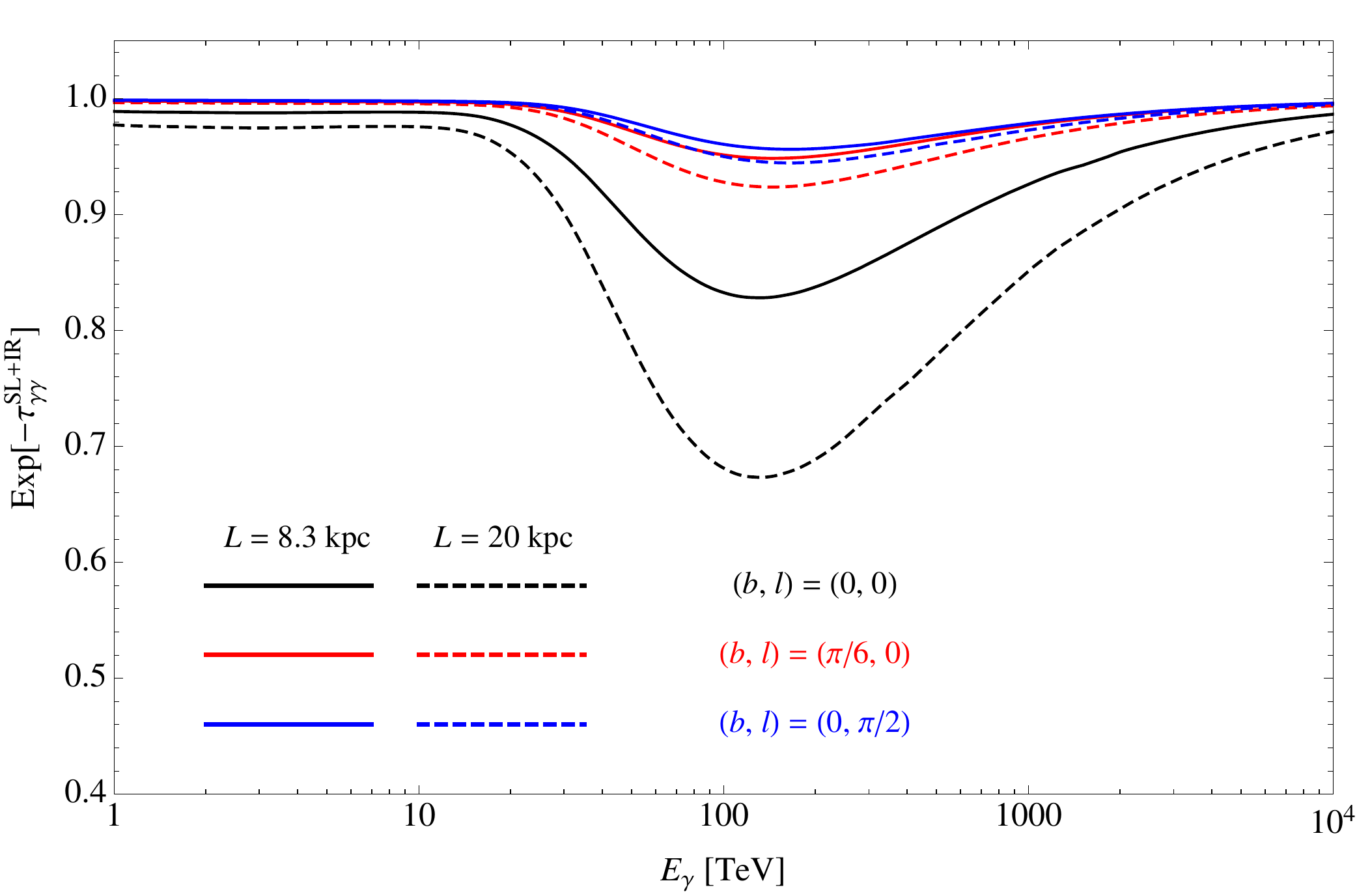}
\caption{Plot of the absorption of $\gamma$-rays on SL+IR photons, for a source at distance $L=8.3$~kpc (solid curves) and 20 kpc (dashed curves) for various directions.}
\label{fig:slirabs}
\end{figure}

The optical depth due to SL+IR photon bath for two different distances and various directions are shown in figure~\ref{fig:slirabs}. It is clear that the absorption effect is relevant around energies of ${\cal O}(100)$~TeV, but only for directions towards the inner Galaxy ($b\simeq l\simeq 0$). The calculated optical depths in this section are consistent with the results reported in~\cite{Moskalenko:2005ng}. The effect of the total opacity of Galactic medium (i.e., $\tau_{\gamma\gamma}=\tau_{\gamma\gamma}^{\rm CMB} + \tau_{\gamma\gamma}^{\rm SL+IR}$) will be discussed in the following sections. 

\section{Expected fluxes from DDM and constraints from EAS experiments}
\label{sec:constraints}

\subsection{Prompt component}
\label{sec:prompt}

\begin{figure}[h]
\centering
\includegraphics[width=0.8\textwidth]{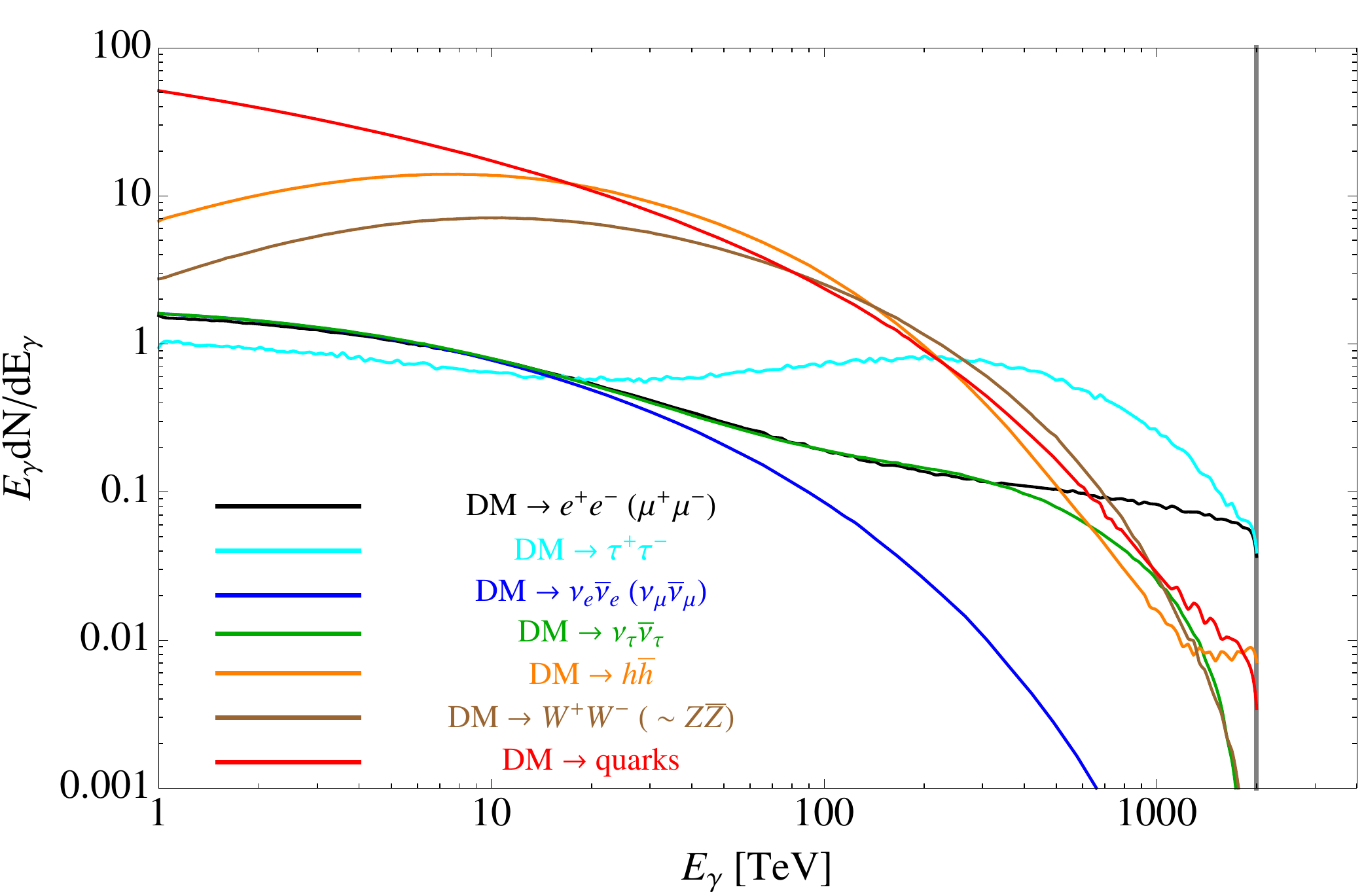}
\label{fig:prompt-lep}
\caption{\label{fig:prompt} Spectrum of the $\gamma$-ray yield in various decay channels of a DM particle with $m_{\rm DM}=4$~PeV. The red curve for DM~$\to$~quarks, shows the average spectra for the DM decay to all the quark flavors. The spectra are obtained via PYTHIA 8.2~\cite{Sjostrand:2014zea}.}
\end{figure}

The prompt component of the Galactic $\gamma$-ray flux from DM decay in the direction $(b,l)$ is given by
\begin{equation}\label{eq:prompt}
\frac{{\rm d} \Phi_{\gamma}}{{\rm d}E_\gamma}(E_\gamma,b,l) = \frac{1}{4\pi\,m_{\rm DM}\,\tau_{\rm DM}} \frac{{\rm d}N_\gamma}{{\rm d}E_\gamma}(E_\gamma) \int_0^\infty \rho_{\rm h}[\varrho(s,b,l)]\, e^{-\tau_{\gamma\gamma}(E_\gamma,s,b,l)}\; {\rm d}s~,
\end{equation}
where $m_{\rm DM}$ and $\tau_{\rm DM}$ are respectively the DM mass and lifetime, and $\tau_{\gamma\gamma}$ is the total optical depth. $ \rho_{\rm h}$ is the density profile of DM particles in our Galaxy as a function of radial distance (in spherical coordinates) from the Galactic center, $\varrho$. For our fiducial model we adopt a Navarro-Frenk-White density profile~\cite{Navarro:1996gj}
\begin{equation}
\rho_{\rm h}(\varrho)\simeq \frac{\rho_h}{\varrho/r_c (1+\varrho/r_c)^2}\;,
\end{equation}
where $r_c\simeq24\,\text{kpc}$ is the critical radius and $\rho_h=0.18\,{\rm GeV}\, {\rm cm}^{-3}$, which yields a DM density at the Solar System $\rho_\odot=0.39\,{\rm GeV}\, {\rm cm}^{-3}$~\cite{Catena:2009mf}. The line-of-sight integration parameter $s$ is related to radial distance $\varrho$ via
\begin{equation}\label{galcoord}
\varrho(s,b,l) = \sqrt{s^2+R^2_\odot-2 s R_\odot \cos b\cos l}\,.
\end{equation}
The $dN_\gamma/dE_\gamma$ is the energy spectrum of photons produced in the decay of a DM particle (here obtained from the PYTHIA 8.2~\cite{Sjostrand:2014zea}, including the weak gauge boson radiation corrections as from~\cite{Christiansen:2014kba}). To illustrate the typical spectra from DM decay, in figure~\ref{fig:prompt} we plot the $E_\gamma dN/dE_\gamma$ for various decay channels of a DM particle with $m_{\rm DM}=4$~PeV. In a specific model of the DM (i.e, specific decay channels with branching ratios determined by the model) the spectrum of $\gamma$-ray can be obtained by the appropriate weighting of the spectra in figure~\ref{fig:prompt}.
In this paper we adopt the scenario introduced in~\cite{Higaki:2014dwa} where the heavy DM particle is a sterile neutrino with mass $\sim 4$~PeV and lifetime $\sim10^{28}$~s, with the branching ratios of the decay channels given by
\begin{equation}\label{eq:br}
{\rm Br}(\ell^\pm W^\mp) = 2\,{\rm Br}(\overset{(-)}{\nu_\ell} Z) = 2\,{\rm Br}(\overset{(-)}{\nu_\ell} h) = \left|U_{\ell1}\right|^2~,
\end{equation} 
where $U_{\ell i}$ are the the elements of the neutrino mixing matrix (for details see~\cite{Higaki:2014dwa}). It is shown in~\cite{Esmaili:2014rma} that this scenario provides a reasonable fit to the energy distribution of IceCube neutrino data. However, let us emphasize that these choices of branching ratios and decay channels are not extremely constrained. In fact, as discussed in~\cite{Esmaili:2013gha}, any model with a sizable  branching ratio into hard (leptonic) channels, with the remaining (even dominant) branching ratio into soft (hadronic and gauge bosons) channels, would provide decent fit to the IceCube data.

\subsection{Including the inverse Compton component}
\label{sec:ICflux}

An additional component of the $\gamma$-ray flux comes from the inverse Compton (IC) scattering of the electrons and positrons from DM decay, up-scattering mostly the CMB photons, which writes 
\begin{equation}\label{eq:IC}
\frac{{\rm d} \Phi_{\rm IC}}{{\rm d}E_\gamma}(E_\gamma,b,l) =\frac{1}{4\pi E_\gamma} \int_0^\infty {\rm d}s\, e^{-\tau_{\gamma\gamma}(E_\gamma,s,b,l)} \int_{m_e}^{m_{\rm DM}/2} {\rm d}E_e  \frac{{\rm d} n_e}{{\rm d}E_e}\left(E_e,\varrho\right)P_{\rm IC}(E_e,E_\gamma,\varrho)~,
\end{equation}
where $\varrho$ is given in eq.~(\ref{galcoord}), $P_{\rm IC}$ is the IC power due to up-scattering on different photon backgrounds and ${\rm d}n_e/{\rm d}E_e$ is the differential number density of $e^+$ plus $e^-$ at steady state. Although the IC flux reported in this article is calculated by taking into account the spatial-dependent nature of energy losses and the effect of spatial diffusion (see appendix~\ref{sec:IC} for the details of the calculation), in order to grasp the main features of IC flux, in the following we pursue a simplified version of the calculation. At the energies of interest here  the transport of electrons and positrons in our Galaxy is determined almost exclusively by the energy losses. Also, one realizes that for directions close to the Galactic plane and for realistic values of the Galactic magnetic field (i.e., $\sim$ few $\mu$G, with a profile that increases towards the inner Galaxy) synchrotron emission is the dominant energy loss mechanism, simply because the synchrotron emission is always quadratic in the electron energy and does not suffer the Klein-Nishina suppression of IC on SL and IR photon baths. Also, at high energies the IC power $P_{\rm IC}$ is almost exclusively due to up-scattering of the CMB photons, and thus independent of the position. The position dependence of the energy loss coefficient, $b=-{\rm d}E_e/{\rm d}t$, is more involved and traces the Galactic magnetic field profile. However, in the approximation in which the thin gaseous disk of the Galaxy is embedded in a thick diffusive halo permeated by a constant magnetic field, the loss coefficient $b$ is independent of the position. In this approximation (which we checked to be accurate whenever the IC signal is non-negligible, see appendix~\ref{sec:IC} for details), one can write
\begin{equation}
\frac{{\rm d} n_e}{{\rm d}E_e}(\varrho,E_e)\simeq \frac{1}{m_{\rm DM}\,\tau_{\rm DM}}\frac{\rho_{\rm h}(\varrho)}{b(E_e)} \frac{{\rm d}N_e}{{\rm d}E_e}(E_e)~,
\end{equation}
and the total $\gamma$-ray flux (i.e., sum of the prompt and IC components) can hence be written as
\begin{equation}\label{halo-tot}
\frac{{\rm d} \Phi_\gamma}{{\rm d}E_\gamma}(E_\gamma,b,l) = \frac{1}{4\pi\,m_{\rm DM}\,\tau_{\rm DM}} 
\int_0^\infty {\rm d}s\; e^{-\tau_{\gamma\gamma}(E_\gamma,s,b,l)}\, \rho_{\rm h}[\varrho(s,b,l)] \left(\frac{{\rm d}N_\gamma}{{\rm d}E_\gamma}+\left.\frac{{\rm d}N_\gamma}{{\rm d}E_\gamma}\right|_{\rm IC}\right)~,
\end{equation}
where
\begin{equation}
\left.\frac{{\rm d}N_\gamma}{{\rm d}E_\gamma}\right|_{\rm IC}(E_\gamma)= \frac{1}{E_\gamma}\int_{m_e}^{m_{\rm DM}/2} {\rm d}E_e \, \frac{1}{b(E_e)}\, P_{\rm IC}(E_e,E_\gamma)\,\frac{{\rm d}N_e}{{\rm d}E_e}(E_e)~.
\end{equation}
${\rm d}N_e/{\rm d}E_e$ is the $e^\pm$ energy spectrum from DM decay, obtained via PYTHIA 8.2~\cite{Sjostrand:2014zea}. $P_{\rm IC}$ can be calculated straightforwardly as reported in appendix~\ref{sec:IC}. Yet, the energy loss coefficient $b$ still depends on the poorly known value of the magnetic field, $B_{\rm halo}$, permeating the thick halo which extends to several kpc away from the disk, and for the lack of better information we approximated it as constant.

\begin{figure}[t] 
\centering
\includegraphics[width=1.\textwidth]{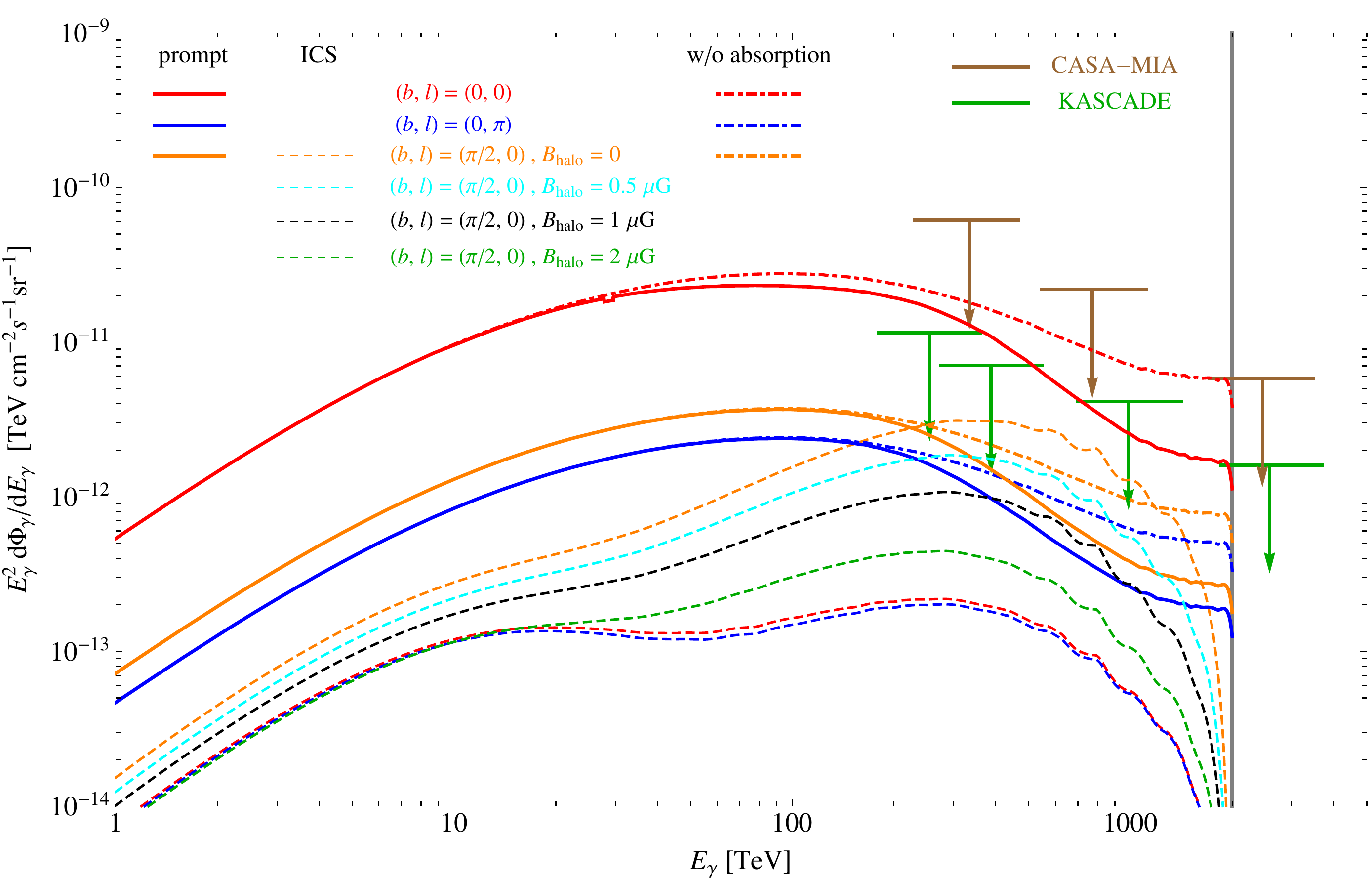} 
\caption{The $\gamma$-ray flux from DM decay from various directions, with $m_{\rm DM}=4$~PeV and $\tau_{\rm DM}=10^{28}$~s, and branching ratios of decay channels given by eq.~(\ref{eq:br}). The solid curves are shows the prompt flux, including the absorption of $\gamma$-rays, while in the dot-dashed curves the absorption is neglected. The dashed curves show the IC flux, for various assumptions for the constant halo magnetic field, $B_{\rm halo}$, possibly pervading the thick diffusive halo of the Galaxy up to large distances. The green and brown bar lines show the upper bound on $\gamma$-ray flux from CASA-MIA~\cite{Chantell:1997gs} and KASCADE~\cite{KASCADE03}, respectively.}
\label{fig:IC}
\end{figure}

In figure~\ref{fig:IC} we show the $\gamma$-ray flux from DM decay, assuming $m_{\rm DM}=4\,$PeV and $\tau_{\rm DM}=10^{28}\,$s (chosen to be close to best-fit parameters; the flux scales inversely with $m_{\rm DM}$ and $\tau_{\rm DM}$) and for the decay pattern with the branching ratios of the decay channels given by eq.~(\ref{eq:br}), for different directions in Galactic coordinates. The solid curves depict the prompt flux, eq.~(\ref{eq:prompt}), from GC (red, top), anti-GC (blue, bottom) and Galactic Pole (orange, intermediate). In each of these curves the dot-dashed curve deviating from the solid curve at higher energies shows the flux neglecting the absorption of $\gamma$-rays discussed in section~\ref{sec:absorption}. When comparing the expected $\gamma$-ray flux from DDM with the experimental bounds, the importance of accounting properly for  the absorption of $\gamma$-ray on CMB photons is manifest, particularly at high energy. The dashed curves in figure~\ref{fig:IC} show the IC flux: the red (blue) dashed curves are the IC flux form GC (anti-GC) direction. The orange dashed curve shows the IC flux from the Galactic pole direction, with the assumption that the Galactic magnetic field only consists of the (thin disk) regular field, given in eq.~(\ref{eq:mag}); i.e., $B_{\rm halo}=0$. The cyan, black and green dashed curves show the IC flux from the Galactic pole within the assumptions $B_{\rm halo}=0.5$, 1 and 2 $\mu$G, respectively. Finally, the green and brown bar lines with arrows show, respectively, the upper limits on the $\gamma$-ray flux inferred by CASA-MIA~\cite{Chantell:1997gs} and KASCADE~\cite{KASCADE03} experiments.

Let us elaborate on the various IC fluxes shown in figure~\ref{fig:IC}: the IC component is clearly sub-leading with respect to the prompt emission for directions along the Galactic plane (note the red and blue dashed curves). However, this is not necessarily the case for the IC flux from the Galactic poles. The enhancement of the IC flux from the Galactic pole direction originates from the fact that for vertical directions the $b$ coefficient drops fasters than the DM density along the line of sight integration of eq.~(\ref{halo-tot}). This enhancement is sizable if one assumes that the magnetic field exponentially decreases for vertical directions (with the profile of eq.~(\ref{eq:mag}))---dashed orange curve in figure~\ref{fig:IC}---so that the IC flux can become comparable to the prompt flux towards the Galactic pole. However, as we have mentioned earlier, it is realistic to assume that a non-zero magnetic field permeates a thick halo to large distances, as consistent with the assumption that a charged cosmic ray population still propagate diffusively in a region several kpc away from the disk. The constant $B_{\rm halo}$ is a toy-model representation of this field, and its effect on the IC flux can be seen by the cyan, black and green dashed curves. In all cases, the emission is suppressed with respect to the ``unmagnetized halo'' situation. The reason is that a growing $B_{\rm halo}$ leads to a larger energy loss coefficient $b$, and thus more suppressed IC flux, since a growing fraction of energy is channeled into synchrotron. In conclusion, the IC flux from directions close to the Galactic plane (low-$b$) is quite robustly predicted to be small. The exact value of IC flux towards the Galactic poles is hard to predict due to the uncertain thickness and $B$-field strength of the magnetized halo, with the orange dashed curves in figure~\ref{fig:IC} providing
a reasonable upper limit to this uncertain component.    

\begin{figure}[h]
\centering
\subfloat[CASA-MIA]{
\includegraphics[width=0.9\textwidth]{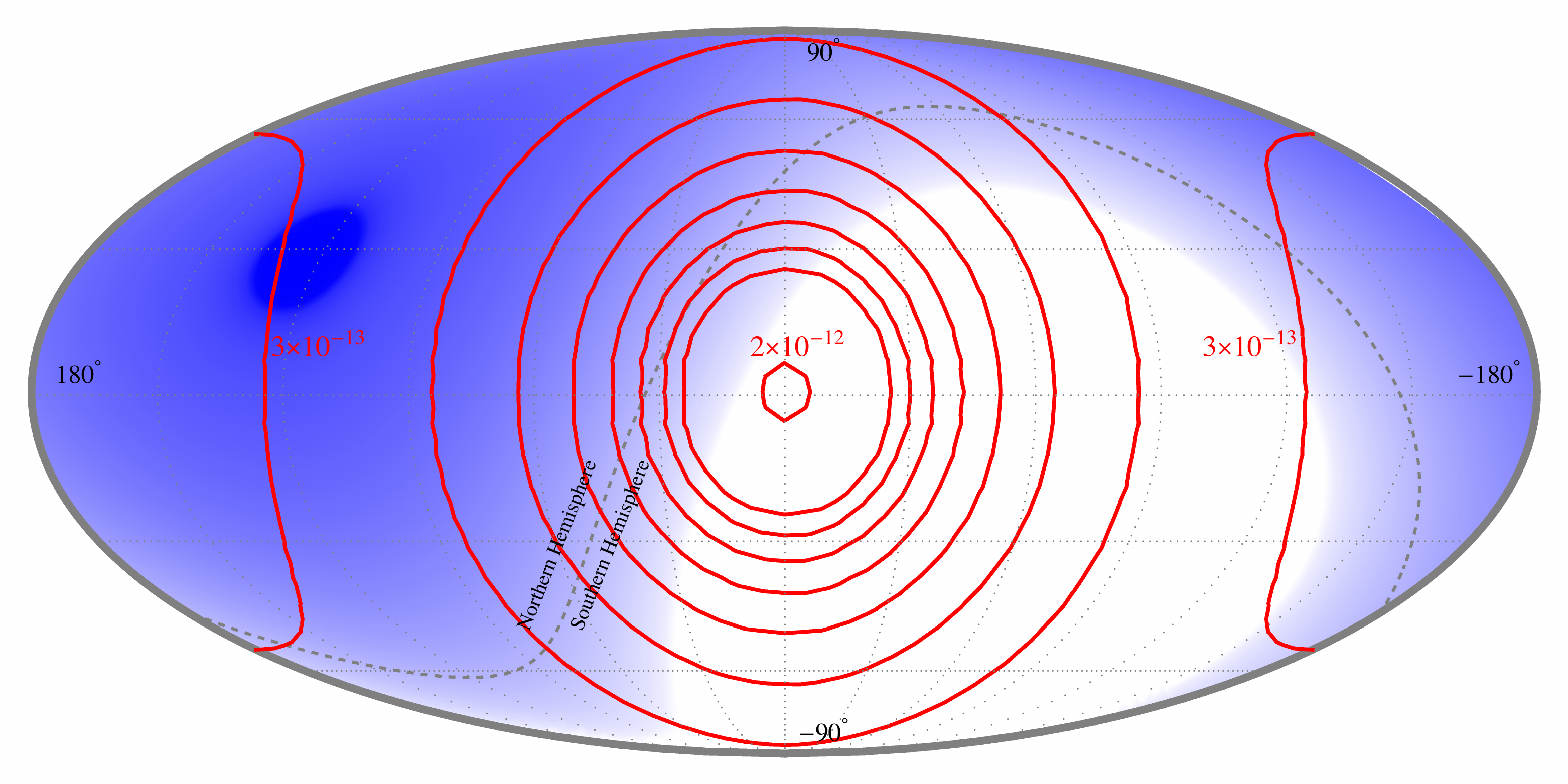}
\label{fig:mol-casa}
}\\
\subfloat[KASCADE]{
\includegraphics[width=0.9\textwidth]{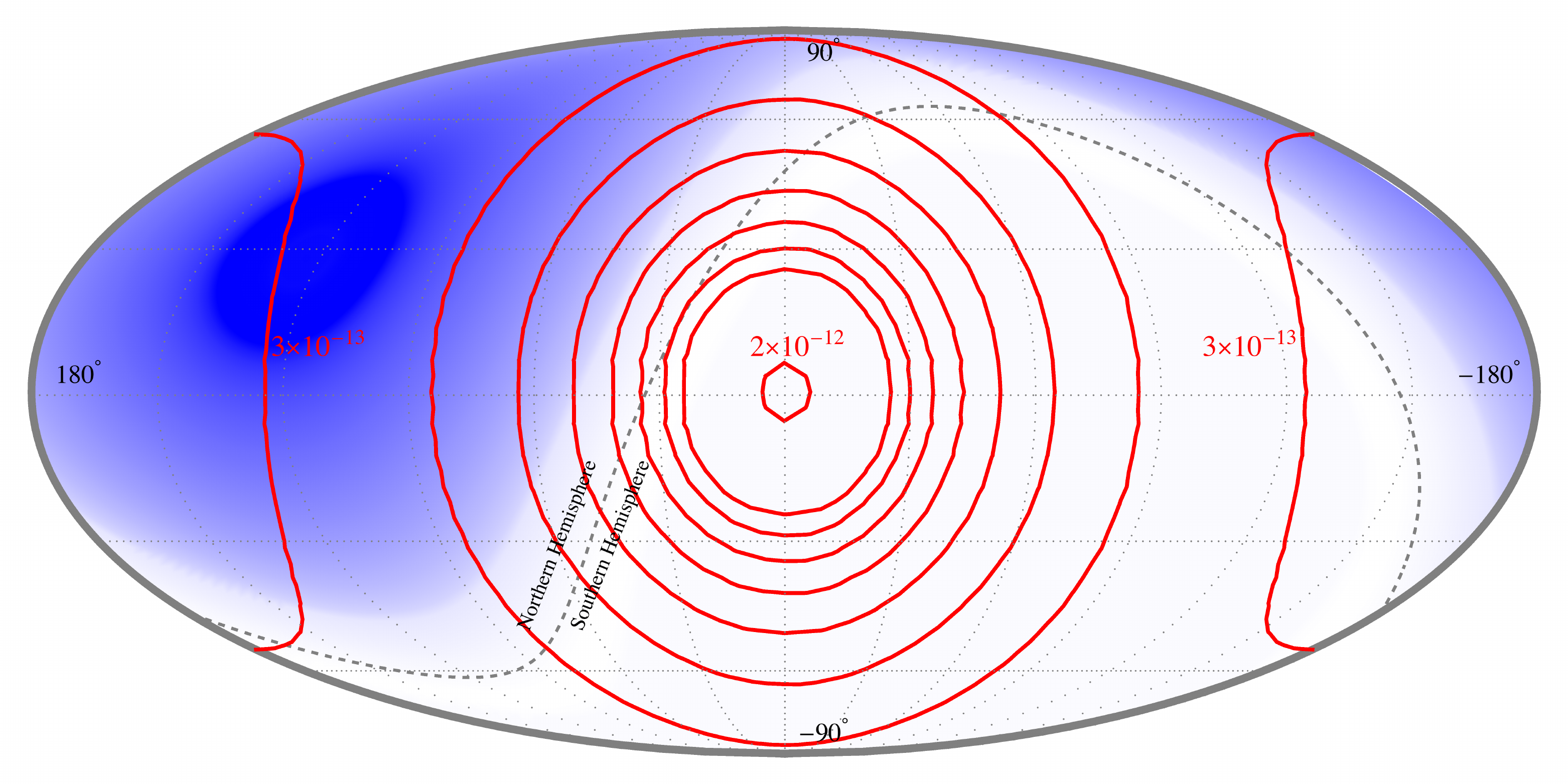}
\label{fig:mol-kas}
}
\caption{\label{fig:mol}The Mollweide plot of the efficiency of EAS experiments (a) CASA-MIA and (b) KASCADE, in Galactic coordinates, discussed in eq.~(\ref{eq:accept}). The red curves show the contours of $E_\gamma^2{\rm d}\Phi_\gamma/{\rm d}E_\gamma$ at $E_\gamma=1$~PeV, with values $2\times10^{-12}$ down to $3\times10^{-13}$, respectively from the inner to outer circles (in the unit TeV cm$^{-2}$ s$^{-1}$ sr$^{-1}$), from the decay of DM with $m_{\rm DM}=4$~PeV, $\tau_{\rm DM}=10^{28}$~s and branching ratios given in eq.~(\ref{eq:br}).}
\end{figure}

It is worth noting that the CASA-MIA and KASCADE experiments would have already probed interesting parameter space for DM models, if they had accumulated significant exposure towards inner Galaxy, e.g.~if they had been located in the Southern hemisphere. Unfortunately, their acceptance mostly peaks in regions far away from the GC and hence they would have been exposed to more modest fluxes, comparable to the orange curve in figure~\ref{fig:IC}, insufficient to test the model even for optimistic IC expectations. To illustrate this point, in the following we briefly describe some notions on the geometrical acceptance of EAS experiments. An EAS is often classified as  $\gamma$-like event, as opposed to a hadronic-like event, based on a significantly poorer  muon content of the former shower with respect to the latter (at a fixed primary energy). Only for events which are not too inclined with respect to the vertical this separation can be done meaningfully, thus imposing a cut on maximum zenith angle of the shower. Assuming that the detector is continuously operational (i.e., the acceptance is uniform with respect to azimuth, or right ascension in equatorial coordinate), the geometrical acceptance efficiency $\omega$ of an EAS experiment located at the latitude $\lambda$ as function of declination $\delta$, can be written as~\cite{Sommers:2000us}   
\begin{equation}\label{eq:accept}
\omega(\delta) \propto \cos\lambda \cos\delta \sin\alpha_m + \alpha_m \sin\lambda \sin\delta~,
\end{equation}        
where 
\begin{displaymath}
\alpha_m = \left\{
  \begin{array}{lc}
       0 & \quad \zeta>1\\
       \pi & \quad\zeta <-1 \\
       \arccos(\zeta) & \quad -1 < \zeta < 1
  \end{array}\right. \qquad {\rm and} \qquad \zeta=\frac{\cos\theta_m - \sin\lambda\sin\delta}{\cos\lambda\cos\delta}~,
\end{displaymath}
where $\theta_m$ is the maximum zenith angle acceptance of the detector. For the CASA-MIA experiment $(\lambda,\theta_m) = (40.2^\circ,60^\circ)$~\cite{Chantell:1997gs}, and for KASCADE $(\lambda,\theta_m) = (49^\circ,35^\circ)$~\cite{KASCADE03}. The blue shaded regions in figures~\ref{fig:mol-casa} and \ref{fig:mol-kas} show the estimated acceptance of CASA-MIA and KASCADE experiments, respectively, in Galactic coordinates. Darker blue regions correspond to higher effective exposure to a $\gamma$-ray flux. The superimposed red curves on the plots are the contours of $E_\gamma^2{\rm d}\Phi_\gamma/{\rm d}E_\gamma$ at $E_\gamma=1$~PeV, with values $2\times10^{-12}$ down to $3\times10^{-13}$, respectively from the inner to outer circles (in units TeV cm$^{-2}$ s$^{-1}$ sr$^{-1}$), from the decay of DM with $m_{\rm DM}=4$~PeV, $\tau_{\rm DM}=10^{28}$~s and branching ratios given in eq.~(\ref{eq:br}). From figure~\ref{fig:IC}, it follows that the upper limits on $E_\gamma^2{\rm d}\Phi_\gamma/{\rm d}E_\gamma$ at $E_\gamma=1$~PeV, are $\sim2\times10^{-11}$ and $\sim4\times10^{-12}$ respectively from CASA-MIA and KASCADE experiments. However, these upper limits apply to the dark blue regions of figure~\ref{fig:mol} and, as can be seen, the limits relax by moving to the regions where the $\gamma$-ray flux from DM increases. In fact in the regions where these experiments are mostly sensitive, the expected flux from decaying DM is $\sim3\times10^{-13}$~TeV cm$^{-2}$ s$^{-1}$ sr$^{-1}$, which is almost one order of magnitude below the KASCADE upper limit.

\section{Anisotropy}
\label{sec:aniso}

\begin{figure}[!t] 
\centering
\includegraphics[width=0.8\textwidth]{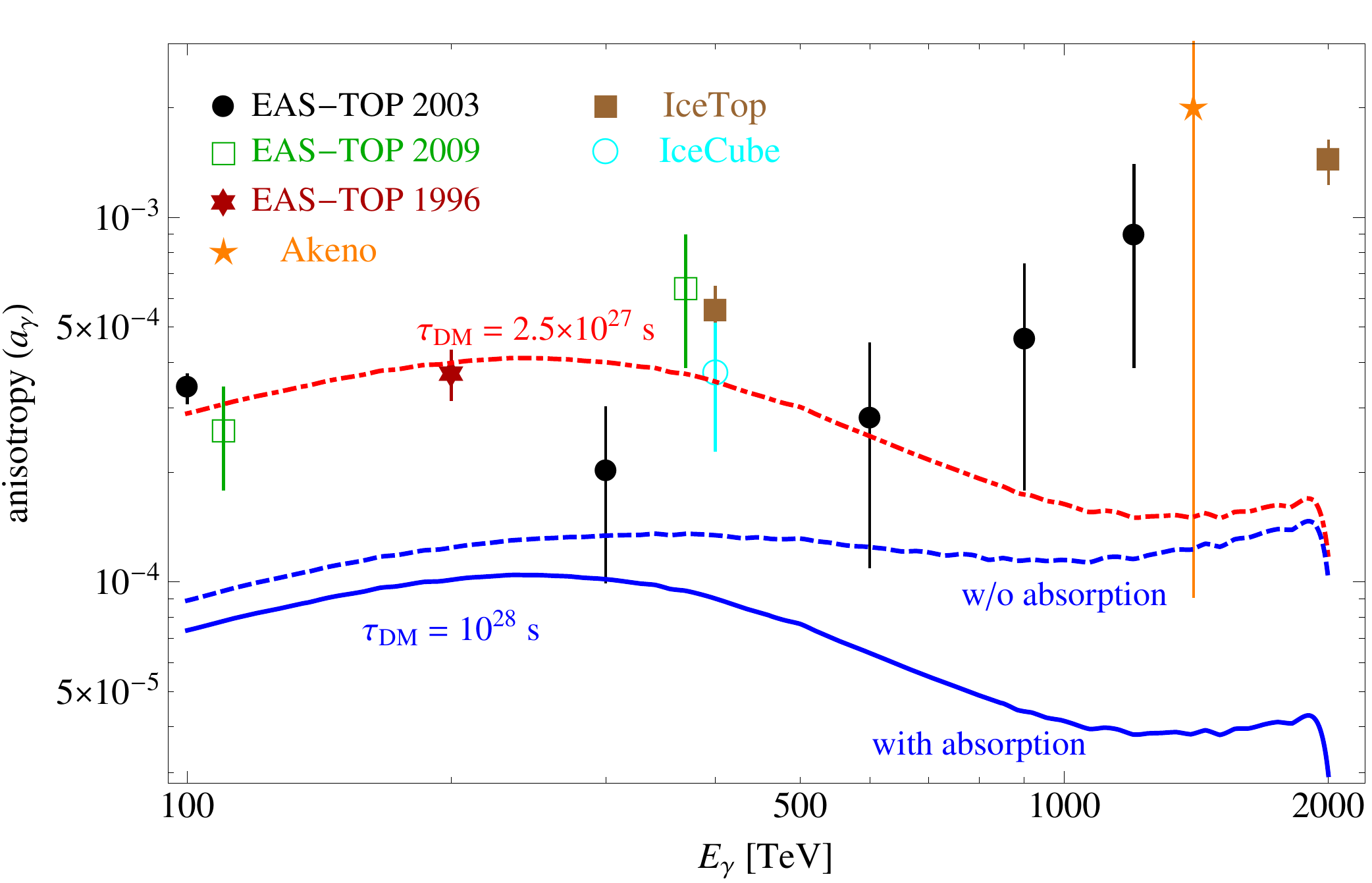} 
\caption{The dipolar anisotropy induces by the $\gamma$-rays, as defined in eq.~(\ref{eq:aniso}), as function of energy. The blue solid (dashed) curve depict the anisotropy by taking into account (neglecting) the absorption of $\gamma$-rays, for $\tau_{\rm DM}=10^{28}$~s. The red dot-dashed curve shows the anisotropy for $\tau_{\rm DM}=2.5\times10^{27}$~s which is the lower limit on lifetime at $2\sigma$ from anisotropy data. The data points show the measured anisotropies by EAS-TOP~\cite{Aglietta:1996sz,Kajita:2003ux,Aglietta:2009mu}, Akeno~\cite{Kifune:1985vq}, IceTop~\cite{Aartsen:2012ma} and IceCube~\cite{Abbasi:2011zka} experiments.}
\label{fig:anisotropy}
\end{figure}

Despite the fact that current EAS bounds are not yet constraining enough for the DDM explanations of IceCube events, the interesting parameter space appears within reach. Even relatively modest optimizations of current sensitivities might thus prove crucial. In fact, the main reason for the degradation of the bounds discussed in the previous section relates to the incorrect assumption that the gamma-ray flux is isotropic. In this section, we discuss to what extent one may turn that weakness into an opportunity, suggesting that {\it anisotropy studies alone}, even without shower property discrimination capabilities, might contribute to the constraints. EAS experiments in fact routinely measure cosmic ray anisotropy, albeit often only in terms of some ``partial estimator'' like the dipolar anisotropy (averaged with respect to right ascension). Let us define a characteristic ``gamma-ray induced anisotropy'' as 
\begin{equation}\label{eq:aniso}
a_\gamma = \frac{\left. \frac{{\rm d}\Phi_\gamma}{{\rm d}E_\gamma}\right|_{\rm GC}-\left. \frac{{\rm d}\Phi_\gamma}{{\rm d}E_\gamma}\right|_{\rm anti-GC}}{\frac{{\rm d}\Phi_{\rm CR}}{{\rm d}E}}~,
\end{equation}
where ${\rm d}\Phi_{\rm CR}/{\rm d}E$ is the total cosmic ray flux, taken from~\cite{Antoni:2005wq}. The anisotropy variable as defined in eq.~(\ref{eq:aniso}) mainly arises from the prompt flux and the contribution of IC flux is negligible, not only because the IC is sub-leading but also since it is expected to be relatively more isotropic. An immediate constraint on DM lifetime can be obtained by requiring that $a_\gamma$ does  not exceed the observed total anisotropy in cosmic rays, $a$. In practice, by requiring that in no energy bin $a_\gamma$ exceeds by more than two sigma the measured value of $a$ we can obtain a conservative bound on the DM lifetime as $\tau_{\rm DM} > 2.5\times10^{27}$~s. The power of this observable is due to the fact that the intrinsic anisotropy in charged cosmic rays is at the level of $10^{-4}\div 10^{-3}$, while a much larger (by two to three orders of magnitude!) relative anisotropy in gamma-rays is expected, at very least due to the off-center position of the Sun in the Galaxy. This means that, despite the fact that gamma-rays only constitute a small fraction of the overall CR flux at $0.1-1$ PeV energies, in the anisotropy observable one can benefit from a larger signal to noise ratio. Accounting for absorption, however, suppresses the gamma-ray anisotropy, since pair-production is more severe in the GC direction than the anti-GC direction. In figure~\ref{fig:anisotropy} the blue solid (dashed) curve shows the expected anisotropy $a_\gamma$ (without) taking into account the absorption, for the fiducial choice of lifetime discussed previously; while the red dot-dashed curve corresponds to the limiting value when $a_\gamma$ exceeds the measured $a$ at $2\sigma$. For comparison, we also report the amplitudes of dipolar anisotropies measured by different experiments. A few remarks are in order:
\begin{itemize}
\item  The suppression of anisotropy due to the absorption can be clearly seen.  It also contributes to the peculiar energy dependence of $a_\gamma$, decreasing with energy, while the observed anisotropy $a$ moderately increases with energy.
\item perhaps surprisingly, the bounds following from anisotropy are {\it at least} comparable in strength with the previously obtained bounds coming from comparisons with the (prompt) flux limits from EAS detectors  and Fermi-LAT diffuse isotropic data, at the level of $10^{27}\,$s.
\item The $a_\gamma$ observable, on the other hand, has a higher sensitivity to the inner Galaxy DM profile. For instance, the previously quoted bound of $2.5\times10^{27}\,$s for the fiducial NFW profile would degrade to $1.9\times10^{26}\,$s for a cored isothermal profile~\cite{Begeman:1991iy} with $\rho_{\rm h}(\varrho)=\rho_h/(1+(\varrho/r_c)^2)$ where $r_c=4.38\,\text{kpc}$ and $\rho_h=1.387\,{\rm GeV}\, {\rm cm}^{-3}$.
\end{itemize}

One may wonder to what extent the above considerations are spoiled by a realistic account of experimental angular and energy resolutions. For instance, CASA-MIA had an angular resolution going from $\gtrsim 2^\circ$ at low energies to better than 0.4$^\circ$ at high-energies~\cite{CASAMIAresA}. This is almost irrelevant for a cored DM profile, while a ${\cal O}(1^\circ)$ resolution can degrade the bound for a NFW profile down to $6\times10^{26}\,$s. Nonetheless, this constitutes a sub-leading uncertainty with respect to the one coming from the unknown shape of the DM profile in the inner Galaxy. Concerning energy resolution, it was demonstrated that CASA-MIA could detect spectral features not larger than $0.2-0.3$ dex in energy~\cite{CASAMIAresE}. Recent cross-calibrations between the energy scale retrieved via surface detectors and the one inferred via fluorescence light suggest that there might be an over-estimate on the absolute energy scale of the former experiments of the order of 30\%, see for instance~\cite{Engel07}. Fortunately, despite these uncertainties, figure~\ref{fig:anisotropy} shows that the expected signal peak is very broad, changing by no more than $\sim 20\%$ between 150~TeV and 350~TeV. We do not expect thus that accounting for energy resolution and energy scale uncertainty would degrade our conclusions by more than  ${\cal O}(10\%)$. It is also interesting that at few hundreds TeV the expected anisotropy from DDM matching IceCube observations may be only $\sim$ one order of magnitude below the measured overall dipolar anisotropy, while at higher energies ($\sim$~PeV) the suppressed anisotropy is smaller by a factor of few, and its ratio to the charged cosmic ray signal is significantly less favorable. This suggests a first potential strategy to improve the constraints by using the energy-dependence and phase information of the anisotropy: although no deterministic prediction of the expected anisotropy due to charged cosmic rays is possible,  one could calibrate a model for the stochastic  phase fluctuation on the high-energy bins (say, 700 to 2000 TeV) where the charged cosmic ray contribution to $a$ is definitely expected to dominate, and use the low energy band (around 200-300 TeV), where the fractional contribution of DM is expected to be maximal, to put a constraint on the amplitude of a sub-leading contribution to the total anisotropy due to $a_\gamma$. The latter is characterized by a fixed direction (constant phase) and a specific energy dependence, very different from the competing charged cosmic ray anisotropy. 

Despite some room for improvement, with past data one cannot really expect order-of-magnitude gains in sensitivity. However, a yet more powerful way to improve over the present analysis would be to rely on the gamma/hadron discrimination possibility attained by the present generation of EAS detectors. In general, cuts based on the morphology of the shower (sometimes dubbed ``compactness'' criteria) allow one to select a photon-rich sample, keeping $\epsilon_\gamma$ fraction of the initial photons, while retaining only $\epsilon_h\ll \epsilon_\gamma$ of the contaminating hadronic background. The ratio $Q\equiv\epsilon_\gamma/\sqrt{\epsilon_h}$ allows to quantify the gain in sensitivity to a photon signal when this cut is applied. While some rejection capability was already present in old experiments, even gamma-ray astrophysics oriented EAS detectors of the past generation, such as MILAGRO, were limited by $\epsilon_h\simeq 0.05-0.1$~\cite{HAWCwebsite} with corresponding $Q$-factors never much larger than $\sim 2$. On the other hand, the situation is significantly different already at currently operating water Cherenkov EAS gamma observatories, such as HAWC. Such an experiment has similar energy resolution performances as the above-mentioned ones (about $40\%$ at $E_\gamma>10$~TeV~\cite{HAWCwebsite}), and even better angular resolution, about $0.1^\circ$ at $E_\gamma>10$~TeV~\cite{HAWCwebsite}. But the major improvement is in the  rejection capability of the hadron background: at high-energies, 99.9\% of the background is routinely rejected, but stringent cuts with $\epsilon_h\simeq 10^{-4}$ and $\epsilon_\gamma\simeq25\%$  above 10 TeV, i.e. $Q\simeq 30$, have already been reported~\cite{Pretz15}, with even better performances that could be attained~\cite{Smith15}. With an effective area, $A_{\rm eff}$, approaching $\sim10^5\,$m$^2$ at high energy and a field of view of $\Delta\Omega\sim2$ sr, HAWC is expected to reach a sensitivity below the level of the IceCube diffuse neutrino flux, thus providing a unique constraint on the electromagnetic counterpart of the neutrino signal~\cite{Pretz15}. A high-energy photon-enriched sample in $T=1$ year of HAWC data would consist of about
\begin{equation}
\epsilon_\gamma \,T\Delta\Omega \,A_{\rm eff} \,\Phi_\gamma \left(E_\gamma>10\,{\rm TeV}\right) \simeq 10^5~{\rm events}~, 
\end{equation}
against a number of background CR events $\epsilon_h \,T\Delta\Omega \,A_{\rm eff} \,\Phi_{\rm CR} \left(E_{\rm CR}>10\,{\rm TeV}\right) \simeq 3\times10^6$. This would certainly ease the measurement of the gamma-ray spectrum, as already noted in the past~\cite{Ahlers:2013xia}. However, it is perhaps even more remarkable that the expected anisotropies of ${\cal O}(10\%)$ in the gamma-ray sample correspond to variations in gamma-ray counts of  $\sim 10^4$, as opposed to anisotropies in the CR background which are expected to be of $\sim 10^3$ events. Put otherwise, in the gamma enriched sample the {\it anisotropy should be fully dominated by the gamma-ray contribution} and, with such statistics, HAWC may provide a crucial test of the DM hypothesis through {\it anisotropy} studies, besides spectral ones.

The situation may be even more favorable with future detectors. HiSCORE~\cite{HISCORE} has two orders of magnitude larger effective area than HAWC at high-energy, but $1\lesssim Q\lesssim 2$ and thus is not ideal for this kind of measurement, although it may still be useful for complementary studies~\cite{Ahlers:2013xia}. However LHAASO~\cite{LHAASOweb}, thanks to its optimized hadron rejection capability, would provide the ultimate sensitivity for this type of analysis: According to~\cite{LHAASO}, thanks to the KM2A array (for detecting hadronic induced muons) surrounding the $10^5\,{\rm m}^2$ Cherenkov detector, at $E_\gamma> 80\,$TeV LHAASO would reach $\epsilon_\gamma\simeq 1$ and $\epsilon_h\lesssim 10^{-7}$, which ensures that observations would be essentially CR background-free.

In summary, anisotropy data offer a complementary tool to constrain DDM contribution to the gamma-ray flux at $0.1-2$~PeV energy. A simple analysis shows that current constraints from the normalization of the anisotropy are competitive with the other methods, and we sketched two possible approaches to improve upon them: with a reanalysis of current data, the addition of phase information could already allow one to achieve sensitivity to a sub-leading DM induced anisotropy. The optimal energy window for DM signal to CR noise appears to be at $\sim200-300$~TeV, while higher energies should be dominated by the CR component, and could be used to calibrate the dominating background anisotropy, whose phase is fluctuating with energy. A major improvement relies however on the current (HAWC) and forthcoming (LHAASO) generation of EAS gamma-ray detectors: thanks to their greatly enhanced photon/hadron rejection capabilities, already in HAWC the anisotropy signal should be dominated by the gamma-ray one, allowing for a first stringent test of any ``Galactic based'' model for the origin of a significant fraction of IceCube events. In LHAASO, we expect the sample to be essentially background-free, allowing for the ultimate test (or detailed studies, in case of positive detection) of these models. Needless to say, these experiments have a great potential for heavy dark matter constraints, which has only recently started to  be studied.

\section{Conclusions}
\label{sec:conc}

The IceCube discovery of a PeV flux of astrophysical neutrinos has several implications for high energy astrophysics and astroparticle physics, as proven by the broad community whose attention has already triggered. If the sources of (a fraction of) these events are Galactic, this discovery paves the way to ``Galactic $\gamma$-ray PeVatrons''. In this regard it is timely to investigate in more detail the peculiarities of VHE $\gamma$-ray propagation in our Galaxy. In fact, the yet undisclosed $0.1-10$~PeV $\gamma$-ray window in Galactic astrophysics would be affected by the pair production absorption. In this article, we discussed some effects that this phenomenon has onto expected signals in extensive air shower (EAS) detectors. We selected the benchmark case of a continuous emission from the Galactic halo in a decaying dark matter (DDM) scenario, although most of our results apply {\it mutatis mutandis} to other source distributions. Our choice was also motivated by the observation that, while the potential of these EAS instruments for DM detection has sometimes been considered in the past, see for instance~\cite{Cuoco:2006tr,Abeysekara:2014ffg}, their potential for serendipitous DM discoveries in more unconventional scenarios has passed mostly unnoticed. 

In this article, we have calculated the expected $\gamma$-ray flux from DDM, with values for the mass and lifetime motivated by the IceCube flux,  by considering both the prompt and IC scattering components. While the angular dependence of the prompt $\gamma$-ray flux is robust (with the detailed features of its energy spectrum depending however on poorly known branching ratios, and only computable within the theoretical uncertainties unavoidably affecting PeV scale physics), the IC scattering component  can potentially exhibit unusual {\it angular} features. Indeed, the IC scattering profile depends on the properties of the magnetized halo of our Galaxy, which is typically poorly known at large vertical distances from Galactic disk. By simple modeling of the Galactic magnetized halo, we have calculated the expected IC flux from directions near Galactic poles, arguing that in the most optimistic case the IC flux can be enhanced, becoming comparable to the prompt flux from this direction. However,  a relatively large halo magnetic field at high latitudes will suppress the IC flux significantly.         

Typical EAS bounds on the $\gamma$-ray fraction in cosmic ray flux have been derived under the hypothesis of a isotropic flux. We argued that this approximation is untenable in the energy range of interest, even in the limit of an isotropically emitted flux, due to the direction (and energy) dependent optical depth of the Galactic sky. We quantified this observation showing that the effect is so relevant that the current constraints from KASCADE or CASA-MIA experiments--which naively would appear to constrain some DDM scenarios--are still at least several times too weak. Ideally, an exposure only marginally better than the above mentioned ones, but at an observatory located in the Southern hemisphere, would be much more promising in this respect. We also argued that anisotropy may offer an independent handle to constrain DDM (as well as other similar scenarios): the expected $\gamma$-ray flux induces an anisotropy in the overall cosmic ray flux only a few times smaller at $\sim\mathcal{O}(100)$~TeV (and about one order of magnitude at $\sim$~PeV) than the current measurements of the dipolar anisotropy routinely performed in EAS experiments. Turning the argument around, existing data are already sensitive to DM lifetimes of ${\cal O}(10^{27})$~s, only one order of magnitude away from the value needed to fit IceCube events ($\sim10^{28}$~s), showing the power of anisotropy analyses and motivating an attempt to improve over the current situation.

Some progress is expected from the  experimental point of view. For instance,  the IceTop facility at the top of the IceCube detector can look at the Southern sky. Unfortunately, while located at the South Pole, the GC is not in standard analyses involving the IceTop array: the IceCube detector plays the role of penetrating muon detector for IceTop facility, which requires that the axis of air shower should pass through the volume of IceCube. This requirement leads to a cut on the zenith angle of shower $\sim30^\circ$ for IC40 configuration~\cite{Aartsen:2012gka}, with the possibility of increasing to $\sim45^\circ$ for the whole IC86 configuration. At the South Pole, the GC is located at zenith angle $\sim61^\circ$. Although the expected sensitivity of IceTop, after 5 years of data taking, is close to the CASA-MIA limit at $\sim$~PeV (see figure 14 at~\cite{Aartsen:2012gka}), due to the closeness of the field of view to GC, IceTop can still moderately improve the limits. Both for past and forthcoming data, we argued that a dedicated analysis might be sensitive to sub-leading anisotropies expected from DDM, notably if the shape of the anisotropy and its energy dependence, whose expectations are relatively well-known, are imposed {\it a priori} in the analysis. 

Eventually, however, we have argued that greatly improved photon/hadron discrimination capabilities are needed for a decisive jump in the sensitivity in both {\it existing} and forthcoming experiments: The recently inaugurated HAWC observatory~\cite{Abeysekara:2014ffg} located in Mexico (with latitude $\lambda\sim19^\circ$ and zenith angle cut $\theta_m\sim45^\circ$) thanks to quality factors $Q=\epsilon_\gamma/\sqrt{\epsilon_h}\sim 30$ could provide first crucial tests of the DDM scenario, not only via spectral studies (see~\cite{Ahlers:2013xia}) but particularly when adding angular information, as discussed in our article. A future experiment such as LHAASO~\cite{LHAASOweb}, benefiting from the KM2A array, is expected to provide a gamma-enriched data set which is almost background (CR)-free, paving the way to exquisite constraints or, in case of detection, to detailed studies of the spectrum and morphology of the signal.

Definitely, the opening of the PeV astrophysical window may offer new opportunities for interesting multi-messenger studies, probably shedding light on intriguing astroparticle questions. As already noticed in the past, and as we further argued here, searches in this new window significantly benefit from EAS experiments. Note that, while the low energy part of the neutrino flux observed by IceCube (recently extended down to $\sim10$~TeV~\cite{Aartsen:2014muf}) can naturally receive contribution from Galactic sources, perhaps easing their identification, pinpointing the origin of the high energy part of the flux is more challenging. In that respect, EAS experiments appear a unique and powerful probe. For such a task, as illustrated in this article, rather than considering the Galactic-scale horizon imposed by the finite optical depth as a limitation, we should perhaps reconsider it as an original opportunity to exploit the specific capabilities of EAS detectors.


\begin{acknowledgments}
We thank M.~Cirelli, P.~Panci, M.~Taoso and F. Vissani for useful discussions, and P. Blasi for discussions and comments on a preliminary version of this manuscript. A.~E. thanks Sebastian Wild for several helps on PYTHIA. At LAPTh, this activity was developed coherently with the research axes supported by the Labex grant ENIGMASS. For A.~E. this research was partially supported by the Munich Institute for Astro- and Particle Physics (MIAPP) of the DFG cluster of excellence ``Origin and Structure of the Universe''. The author would like to express a special thanks to the Mainz Institute for Theoretical Physics (MITP) for its hospitality and support. 

\end{acknowledgments}

\appendix
\section{Details of the IC flux calculation}
\label{sec:IC}

In this appendix we provide the details of the IC flux calculation, which have been used to compute the curves in figure~\ref{fig:IC}. As we discussed, the IC flux is given by eq.~(\ref{eq:IC}) and here we discuss calculation of the ingredients ${\rm d}n_e/{\rm d}E_e$ and $P_{\rm IC}$. 

\begin{figure}[t] 
\centering
\includegraphics[width=0.7\textwidth]{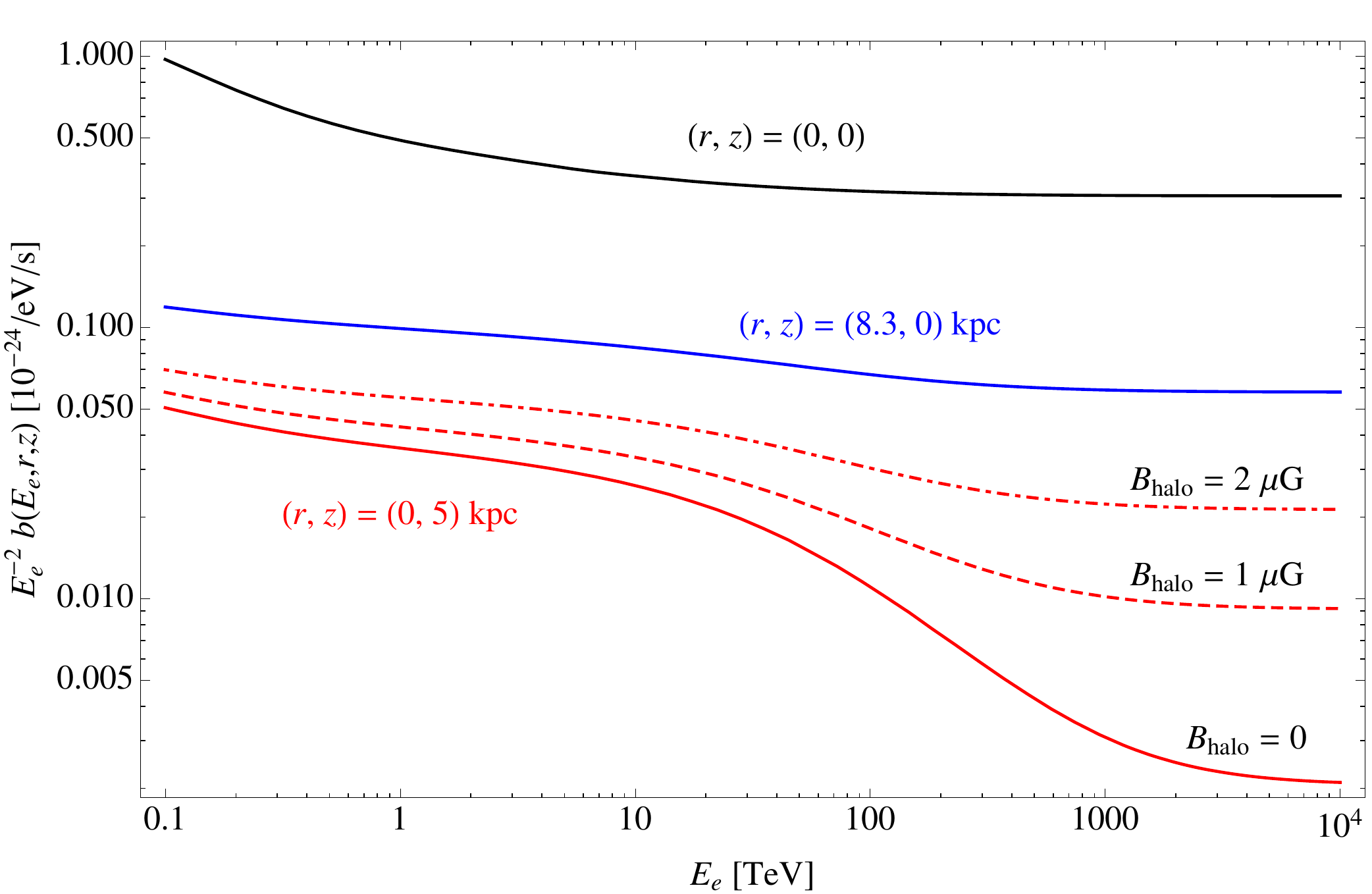}
\caption{Plot of $E_e^{-2} b(E_e,r,z)$ as function of $E_e$ at some positions in the Galaxy. For all curves, the assumed disk regular magnetic field is eq.~(\ref{eq:mag}); while for solid curves $B_{\rm halo}=0$, for dashed red curve $B_{\rm halo}=1~\mu$G and for dot-dashed red curve $B_{\rm halo}=2~\mu$G.}
\label{fig:bplot}
\end{figure}

The steady state spectrum of $e^\pm$ at position $\vec{x}$ in our Galaxy (we will in fact assume azimuthal symmetry in cylindrical coordinates) can be calculated by
\begin{equation}
\frac{{\rm d}n_e}{{\rm d}E_e} (E_e,\vec{x})= \frac{1}{m_{\rm DM}\,\tau_{\rm DM}}\frac{\rho_{\rm h}(\vec{x})}{b(E_e,\vec{x})} \int_{E_e}^{m_{\rm DM}/2} \, \frac{{\rm d}N_e}{{\rm d}E_e^\prime}(E_e^\prime)\, I_{\rm diff} (E_e,E_e^\prime,\vec{x})\, {\rm d}E_e^\prime~,
\end{equation}
where $b$ is the energy loss coefficient and $I_{\rm diff}$ is the function taking into account the diffusion of $e^\pm$. Let us elaborate on each of these ingredients:

\textbf{Energy loss function $b(E_e,\vec{x})$}: The $e^\pm$ lose energy by two main processes: synchrotron radiation in the magnetic field of Galaxy and energy loss due to IC scattering on the photon bath (SL+IR and CMB). Since both the magnetic field and SL+IR field are position dependent, the energy loss function also is position dependent. So,
\begin{equation}
b(E_e,\vec{x}) \equiv - \frac{{\rm d}E_e}{{\rm d}t} = b_{\rm IC} (E_e,\vec{x}) + b_{\rm syn} (E_e,\vec{x})~,
\end{equation}  
where
\begin{equation}
b_{\rm IC} (E_e,\vec{x}) = 3\sigma_T \int_{0}^{\infty} \varepsilon {\rm d}\varepsilon \int _{\frac{1}{4\gamma^2}}^{1} {\rm d}q
 n(\varepsilon,\vec{x}) \frac{(4\gamma^2 - \Gamma_\varepsilon)q-1}{(1+\Gamma_\varepsilon q)^3} \left[ 2q\ln q +q+1-2q^2 + \frac{(\Gamma_\varepsilon q)^2 (1-q)}{2(1+\Gamma_\varepsilon q)}\right]~,  
\end{equation}
where $\sigma_T$ is the Thomson cross section, $\Gamma_\varepsilon = 4\varepsilon\gamma/m_e$ and $\gamma=E_e/m_e$; and $n(\varepsilon,\vec{x})$ is the SL+IR+CMB photons differential number density at position $\vec{x}$. The energy loss due to synchrotron radiation can be calculated by
\begin{equation}
b_{\rm syn} (E_e,\vec{x}) = \frac{4\sigma_T E_e^2 }{3m_e^2}\cdot \frac{B^2(\vec{x})}{2}~,
\end{equation}
where $B$ is the magnitude of the total magnetic field in our Galaxy, consisting of regular and turbulent components~\footnote{The interstellar magnetic field of our Galaxy off the Galactic plane is determined by the Faraday-rotation measurement of the polarization of extragalactic radio sources. However, these measurements depend on the assumed free-electron distribution at high latitudes which is poorly known. On the other hand, even in the Galactic plane, the estimated turbulent magnetic field is a factor of few larger than the regular magnetic field.}. For the regular magnetic field we adopt the following profile~\cite{Strong:1998fr}
\begin{equation}\label{eq:mag}
B_{\rm reg}(\vec{x})\equiv B_{\rm reg}(r,z) = B_0 ~ \exp\left[-\frac{|r-R_\odot|}{r_B} - \frac{|z|}{z_B}\right]~,
\end{equation}  
where $R_\odot = 8.3$~kpc, $r_B=10$~kpc, $z_B=2$~kpc and $B_0 = 4.78~\mu$G. For the halo (possibly turbulent) magnetic field we assume a  uniform constant strength magnetic field. Figure~\ref{fig:bplot} shows the energy loss function $b$, as function of $E_e$, at GC (black solid), Sun position (blue solid) and at vertical distance $z=5$~kpc from Galactic plane (at GC) for three different assumptions for $B_{\rm halo}=0,1$ and 2 $\mu$G, respectively by solid, dashed and dot-dashed red curves. As can be seen by increasing the value of $B_{\rm halo}$ from 0 to 2 $\mu$G, the energy loss coefficient at $z=5$~kpc increases by about one order of magnitude, which justify the suppression of IC flux (black and green dashed curves in figure~\ref{fig:IC}) for larger $B_{\rm halo}$. Obviously, the effect of halo field is smaller at lower energies, since the synchrotron emission is the main mechanism of energy loss at higher energies.
 
\textbf{Diffusion halo function $I_{\rm diff}(E_e,E_e^\prime,\vec{x})$}: The diffusion halo function can be calculated by solving the diffusion-loss equation of $e^\pm$ in the Galaxy. To avoid repetition we skip reporting the details of calculation, which is done according to the prescription reported in~\cite{Cirelli:2010xx}. However, it is worth mentioning that at high energies which we are interested in this paper $I_{\rm diff}\simeq1$, and so the results of of IC flux reported in figure~\ref{fig:IC} are only marginally dependent on the diffusion halo function. Put otherwise, the approximation described in the main text is actually excellent. 

\begin{figure}[h]
\centering
\subfloat[]{
\includegraphics[width=0.5\textwidth]{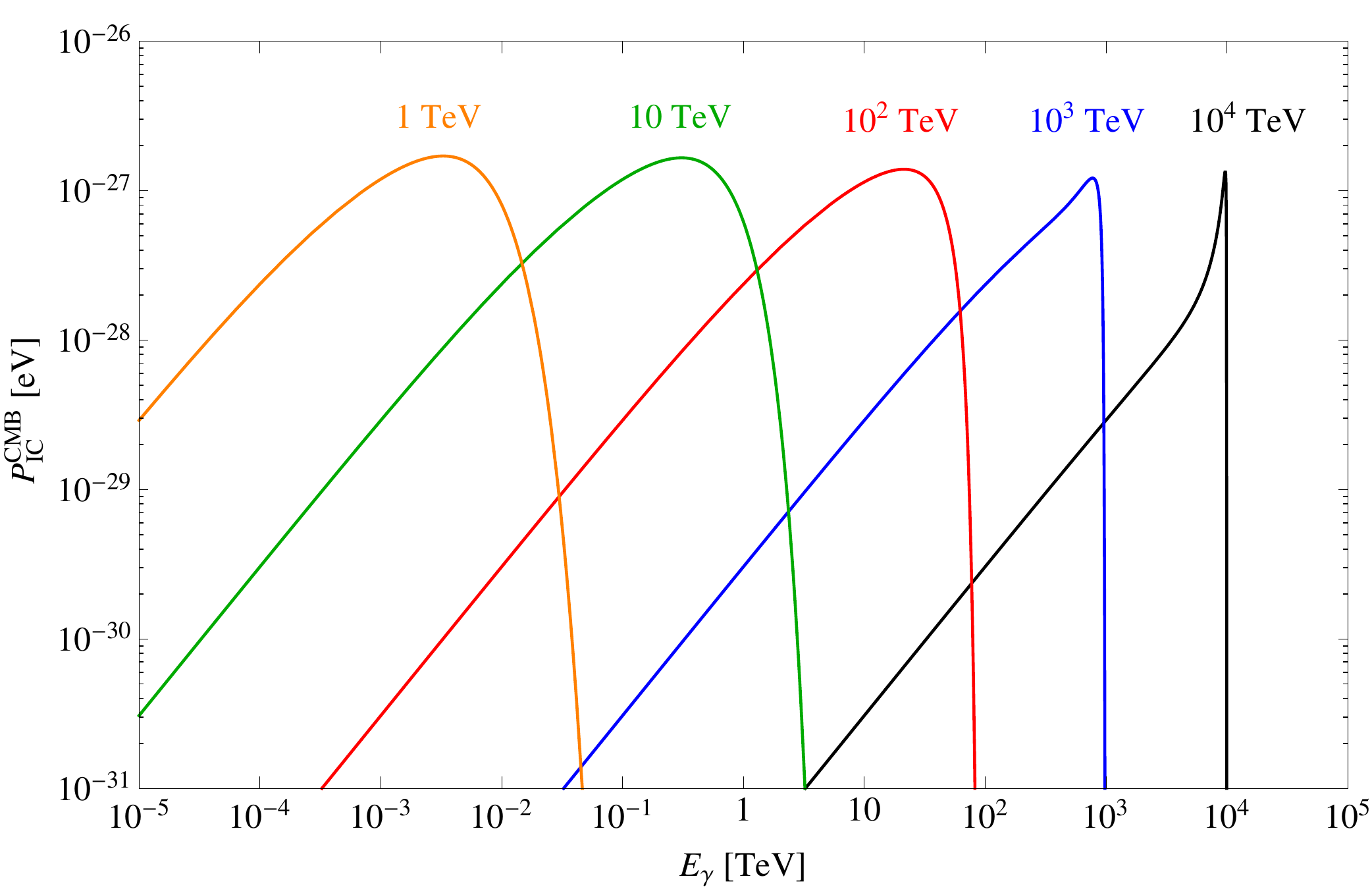}
\label{fig:cmb-pic}
}
\subfloat[]{
\includegraphics[width=0.5\textwidth]{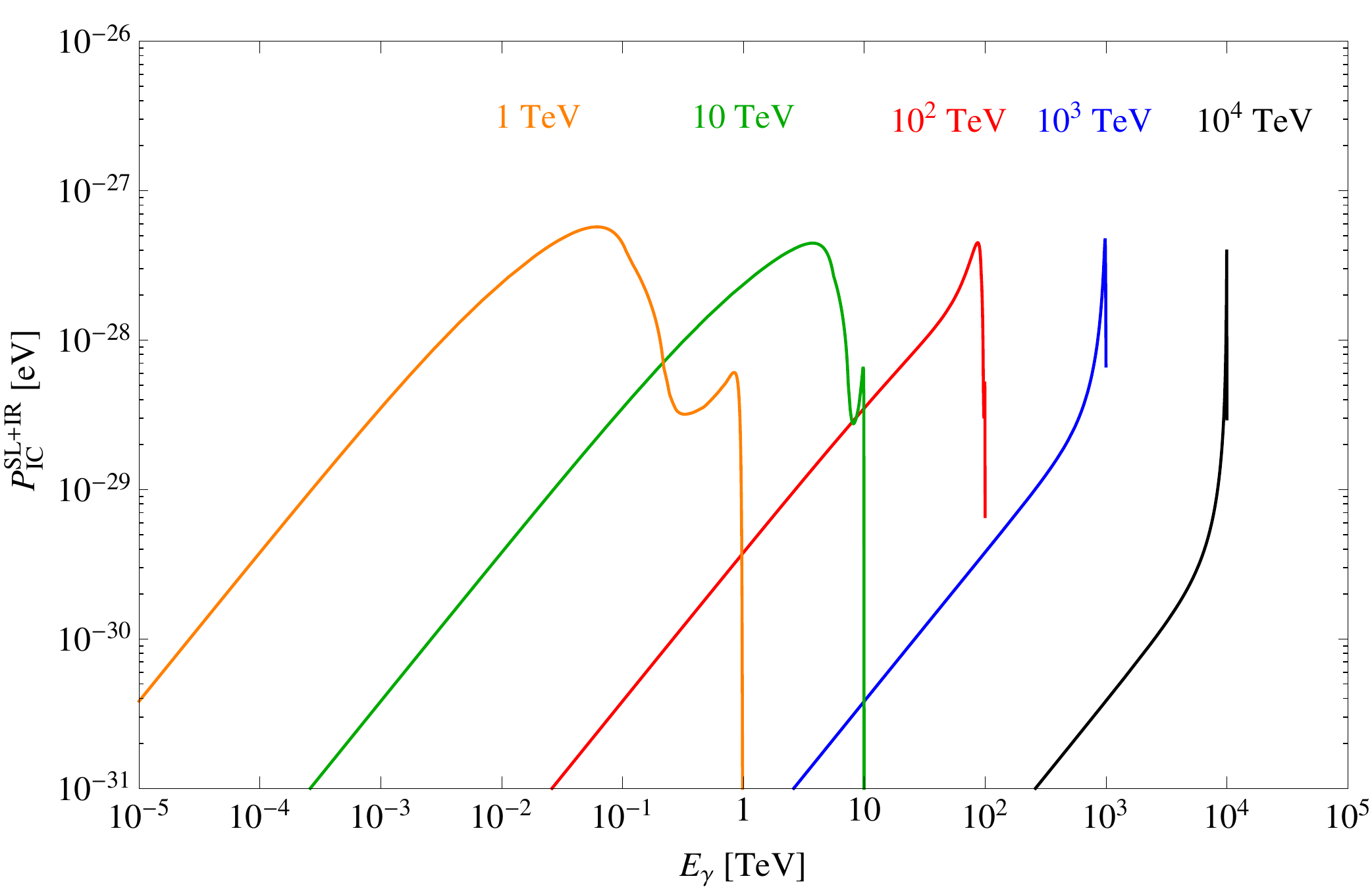}
\label{fig:slir-pic}
}
\caption{\label{fig:pic} Inverse Compton power of: (a) $P_{\rm IC}^{\rm CMB}(E_\gamma,E_e)$, (b) $P_{\rm IC}^{\rm SL+IR}(E_\gamma,E_e,x=0)$, for $E_e=1,10,10^2,10^3,10^4$~TeV, from the left to the right, respectively.}
\end{figure}

The other ingredient in the calculation of IC flux is the IC power $P_{\rm IC}$ (see eq.~(\ref{eq:IC})) which can be decomposed into the IC power from each component of the photon bath; i.e., $P_{\rm IC}=\sum_i P_{\rm IC}^i$, where $P^i_{\rm IC}$ is the IC power from the photon bath $n_i$ with $i=$ CMB and SL+IR. The $P^i_{\rm IC}$ can be written as
\begin{equation}
P_{\rm IC}^i (E_\gamma,E_e,\vec{x}) = \frac{3\sigma_T E_\gamma}{4\gamma^2}  \int_{\frac{1}{4\gamma^2}}^{1} dq~\left[ 1- \frac{1}{4q\gamma^2(1-\epsilon)} \right] \frac{n_i\left( E^0_\gamma(q),\vec{x} \right)}{q} \left[ 2q\ln q +q +1-2q^2+\frac{\epsilon^2(1-q)}{2(1-\epsilon)}\right]~,
\end{equation}
where $\gamma=E_e/m_e$, $\epsilon=E_\gamma/E_e$ and 
\begin{equation}
E^0_\gamma(q) = \frac{m_e^2 E_\gamma}{4qE_e(E_e-E_\gamma)}~.
\end{equation}
Figures~\ref{fig:cmb-pic} and \ref{fig:slir-pic}, respectively, show the IC powers $P_{\rm IC}^{\rm CMB}$ and $P_{\rm IC}^{\rm SL+IR}$ as function of $E_\gamma$ for $E_e=1,10,10^2,10^3,10^4$~TeV at GC. As can be seen, at high energies the IC power sharply peaks at $E_e$. Namely, in the IC scattering almost all the $e^\pm$ energy transfers to the photon. Also, by comparing the corresponding curves in figures~\ref{fig:cmb-pic} and \ref{fig:slir-pic}, it can be seen that at high energies the main contribution to the total IC power comes from $P_{\rm IC}^{\rm CMB}$. The IC power in figure~\ref{fig:cmb-pic} is independent of the position in Galaxy (due to the uniform CMB photon bath), while the IC power due to SL+IR strongly decreases by distancing from the GC, especially in the vertical direction with respect to Galactic plane.

\end{document}